\documentclass[11pt]{article}
\usepackage{subfigure}
\usepackage{parskip}
 \expandafter\let\csname equation*\endcsname\relax
 \expandafter\let\csname endequation*\endcsname\relax
\usepackage{bm}
\usepackage{enumerate}
\usepackage[top=1in, bottom=1in, left=1in, right=1in]{geometry}
\usepackage{mathrsfs,color}
\usepackage{amsfonts}\usepackage{multirow}
\usepackage{amssymb}
\usepackage{amsmath}
\usepackage{float}
\usepackage{graphicx,afterpage}
\usepackage{epstopdf}
\usepackage[pdftex,colorlinks=true]{hyperref}
\usepackage[sort,compress]{cite}

\usepackage{natbib}

\usepackage{upgreek}
\usepackage[dvipsnames]{xcolor}
\usepackage{bm}
\usepackage{algorithm2e} 
\usepackage{hyperref}

\usepackage{multirow}

\usepackage{tikz}

\usepackage{xcolor}
\usepackage{tikz}
\hypersetup{
     colorlinks   = true,
     citecolor     = teal, 
     linkcolor     = teal
}

\newcommand{\eg}{e.g.,~}

\renewcommand{\mod}{\hat{\eta}(x,y,t|\boldsymbol{\theta})}

\renewcommand{\epsilon}{\varepsilon}
\newcommand{\etaobs}{\eta_{\mathrm{obs}}}
\newcommand{\heta}{\hat{\eta}}
\newcommand{\etatest}{\eta_{\mathrm{test}}}
\newcommand{\etatrain}{\eta_{\mathrm{train}}}
\newcommand{\mps}{\mathrm{m}\,\mathrm{s}^{-1}}

\newcommand{\gradeta}{|\nabla \heta|}
\newcommand{\lapeta}{\nabla^2 \heta}
\newcommand{\detadt}{\partial \heta/\partial t}
\newcommand{\detadx}{\partial \eta/\partial x}
\newcommand{\detady}{\partial \eta/\partial y}

\begin{document}
\begin{center}
{\bf\Large{Continuous and Differentiable Neural Network Representations of the Surfzone Sea-Surface from Lidar Point Clouds}}

\vspace{4mm}
David R. Culver, Matthias Morzfeld, Falk Feddersen

\vspace{2mm}
Scripps Institution of Oceanography, University of California, San Diego, CA\\
\end{center}

\begin{center}
\emph{Abstract}
\end{center}
Ocean waves are often analyzed via sea-surface observations.
Multi-beam scanning lidar enables two-dimensional 
measurements of the water surface $\eta(x,y)$ via sequenced point clouds of
sea-surface elevation.  Point cloud data, however, have to be gridded,
interpolated, or mapped for use in wave analysis.  We
present and test a  method for creating a continuous and differentiable
(space and time), but grid-free, representation of the sea-surface from time
sequenced lidar point clouds.  We use a specific neural network architecture, a
SIREN, to represent the sea-surface and to train the network on spatiotemporal
60~s of lidar data collected by a gas-electric UAS (drone) hovering over large (4 m),
breaking waves in Monterey, California.  The trained SIREN allows estimation of
the sea-surface and its derivatives at any location and time. 
SIREN sea-surface representations tend to filter the highest wave number variability (splash-up and sea-spray).
SIREN errors are comparable
to instrument errors and insensitive to local data density.  The reliability and
reproducibility of our approach is tested by training two SIRENs independently
on the same data.  Differences between the two independent SIRENs
are comparable to instrument error, but increase in data-poor regions. 
SIRENs' potential for directional wave analysis is demonstrated by computing
spatial and temporal derivatives and by estimating wave velocity from a
SIREN-generated space-time slice of the sea-surface.

\section{Introduction}

Ocean waves have long been studied via analysis of sea-surface
observations.
For example, surface pitch and roll wave buoys have been used to estimate
wave spectra and wave directional moments at a specific location 
\citep{longuet1963,kuik_method_1988}.
Remote sensing of the sea surface, either by lidar \citep[\eg][]{Melville2016JTECH} or
stereo photogrammetry \citep[\eg][]{bergamasco_wass_2017}, can provide
time-varying point-clouds of the sea-surface.
In the nearshore, where the waves
are steep and changing rapidly, lidar or stereo photogrammetry point clouds of the sea-surface
have been used to visualize a wide range of wave processes from wave steepening, to overturning,
the subsequent splash-up, bore propagation and swash
\citep[\eg][]{brodie_lidar_2015,Odea2021GRL,jaudMeasuringNearshoreWaves2024,Davey2026submit}.
In the nearshore, single beam scanning lidars mounted on shore or on piers have
been used to measure the temporal ($t$) and cross-shore ($x$) varying sea-surface $\eta(x,t)$ across the surf and swash zones
\citep[\eg][]{BLENKINSOPP2010,fiedler_observations_2015,MARTINS2017}.
Single beam scanning lidars were also mounted on uncrewed aircraft system (UAS) with real-time positioning
capability, enabling high grazing angles and better data quality
\citep{Huang2018JTECH,Fiedler2021JTECH}.
Multi-beam scanning lidar allows for the two-dimensional ($x,y$) measurement of the water
surface.   Such lidars have been used to observe  wave overturning
in both field \citep{Odea2021GRL} and field-scale laboratory
\citep{Feddersen2023JFM} settings.
A multi-beam scanning lidar mounted on a UAS was able to reproduce directional
wave statistics from a wave buoy in 10~m water depth \citep{feddersen_estimating_2024}.

For use in analysis of waves, point cloud data must first be gridded, mapped, or fit
to functional forms.   For example, $\eta(x,y,t)$ at discrete resolution has been
estimated from despiked and gridded/interpolated single beam scanning lidar point clouds 
at 0.1~m cross-shore resolution and at a
time resolution defined by the lidar sampling rate 
\citep{brodie_lidar_2015,MARTINS2017,Fiedler2021JTECH}.
Wave overturning geometry was quantified by fitting a subset of point cloud data
to a specific functional form \citep{Odea2021GRL,Feddersen2023JFM}.
Multi-beam lidar point cloud data within a small circular region was fit
to a plane and 2D parabola to estimate $\eta(t)$ and its derivatives $\detadx$
and $\detady$ at specific locations \citep{feddersen_estimating_2024}. 

Various types of wave analysis require estimates of $\eta(x,y,t)$ and its spatial and temporal derivatives.
Nonlinear amplitude
dispersion analysis \citep{Herbers2002JPO,Martins2021JPO}, for example, requires estimating spectra of $\eta$, and its spatial derivatives $\detadx$, and $\detady$. Analysis of the surface kinematic boundary condition requires estimating the temporal derivative $\partial \eta/\partial t$.
We describe how to use a certain type of neural network (NN) as a continuous, grid-free functional representation of the sea-surface $\eta(x,y,t)$, reconstructed from point cloud data (multi-beam lidar), whose value and spatiotemporal derivatives can be estimated anywhere within a space-time domain.

NNs are parameterized functions that can continuously and differentially map inputs to outputs \citep[\eg][]{goodfellow_deep_2016}.
In computer graphics, coordinate based NNs re-render images by evaluating pixel coordinates either as grayscale or RGB, and more sophisticated coordinates can be used to reconstruct 3D shapes~\citep{park_deepsdf_2019,chen_learning_2019}.  
Coordinate based NNs trained to minimize data mismatch 
are known as implicit neural representations (INR), and 
have a  wide range of applications \citep{mildenhall_nerf_2020,tewari_advances_2022}.
INR construction was improved using 
sinusoidal representation networks (SIREN) which
leverage periodic activation functions in the 
NN design~\citep{sitzmann_implicit_2020,novello_understanding_2023,ngom_fourier_2021}.
SIREN representations produce images at higher quality in both the signal and its derivatives \citep{sitzmann_implicit_2020}.
SIRENs can also be used for solving partial differential equations \citep{catalani_neural_2024} or for compressing large video files and point clouds \citep[\eg][]{zhang2021implicitneuralvideocompression, kwan_enhanced_2026,kwan_immersive_2024,rho_neural_2022, zhang_efficient_2025}, and the training process can leverage derivative information as well \citep{yuan_sobolev_2022}.

The continuity and differentiability of SIREN reconstructions offer desirable properties for mapping the sea-surface, especially when compared to grid-based approaches.
Here, we describe in detail how to use a coordinate based SIREN for continuous and differentiable, grid-free representations of the sea-surface $\eta(x,y,t)$, obtained from lidar point cloud data.
The rest of this paper is organized as follows.
We describe the data collection in Section~\ref{sec:Methods}\ref{sec:exp_over}: A multibeam lidar mounted on a UAS was flown over the surf zone of a rocky shore during a large swell-event. 
The design, architecture and training procedure of the SIREN is explained in Section~\ref{sec:Methods}\ref{sec:SIREN_reconstruction}. 
A subset of the data were withheld from the training process, which is then used to study errors in the SIREN representation of $\eta(x,y,t)$ in Section~\ref{sec:Results}\ref{sec:errors}. 
We examine the reliability and reproducibility of SIREN sea-surface representations in Section~\ref{sec:Results}\ref{sec:residuals}. 
In Section~\ref{sec:wave analysis}, we show examples of the SIREN reconstructed $\eta(x,y,t)$ and its derivatives to demonstrate the potential of the new method.
In Section~5, we discuss the advantages and disadvantages of using SIREN representations compared to existing approaches, discuss the effects of different initialization schemes for training, and provide a summary.

\section{Methods}
\label{sec:Methods}

\begin{figure}[tb]
  \centering
    \includegraphics[width =.7 \textwidth]{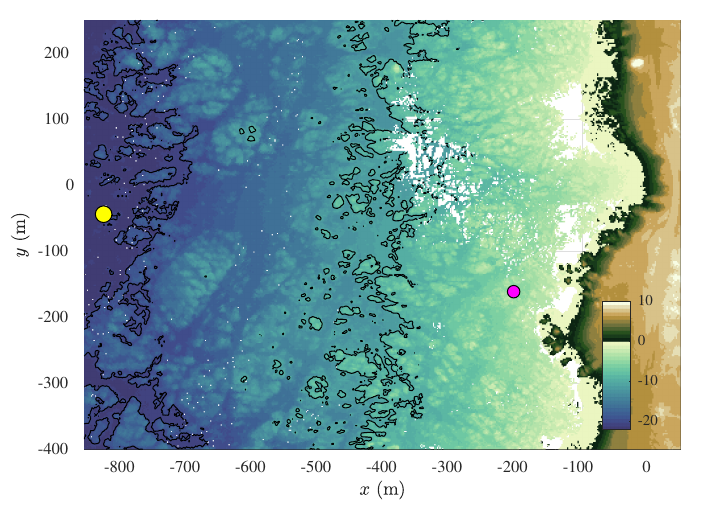}
    \caption{Bathymetry ($z$, meters relative to mean sea level) at the
China Rock region (Monterey Peninsula, CA, USA) as a function of local cross-shore $x$ and along-shore $y$ coordinates.  The origin is located at 36.6044$^\circ$ and -121.9594$^\circ$.
The yellow circle represents the location of the Spotter wave buoy at $(x,y)=(-823,-30)$~m and the
magenta circle indicates the location of the UAS hover $(x,y)=(-203,-161)$~m.
Regions in white indicate locations with no bathymetric observations.}
    \label{fig:map}
\end{figure}

\subsection{Lidar and Experiment Overview}
\label{sec:exp_over}
We use data collected by an eight-rotor Skyfront Perimeter 8 UAS, equipped
with a lidar and an Inertial Navigation System (INS).   
The lidar and UAS were previously reported in \citet{feddersen_estimating_2024}.
A RTK-GNSS module allows the UAS to maintain its position without drifting over time. 
The UAS payload is a Phoenix lidar Systems (PLS) Scout Ultra, which houses a Velodyne Ultra Puck (VLP-32c) lidar, a 
24-MP Sony A6K-Lite RGB camera, and a proprietary INS system made up of an inertial motion unit (IMU) and Novatel OEM7720 GNSS receiver.
The Velodyne Ultra Puck lidar has 32 beams at angle 90$^\circ$ from the nose of the UAS.
Most beams are centrally organized in the vertical field of view, increasing data resolution in the center.
The beams are non-linearly distributed, resulting in a 40$^\circ$ off axis field of view (from -25$^\circ$ to -15$^\circ$). 
The beams rotate 360$^\circ$ at 10~Hz, providing sea-surface data at 10~Hz.
Raw lidar returns are put into earth coordinates using the post processed position and
orientation solutions. 
All points closer than 8~m and farther than 100~m from the lidar were removed as quality control.
Full details of the UAS and lidar are found in \citet{feddersen_estimating_2024}.

\begin{figure}[tb]
    \centering
    \includegraphics[width = \textwidth]{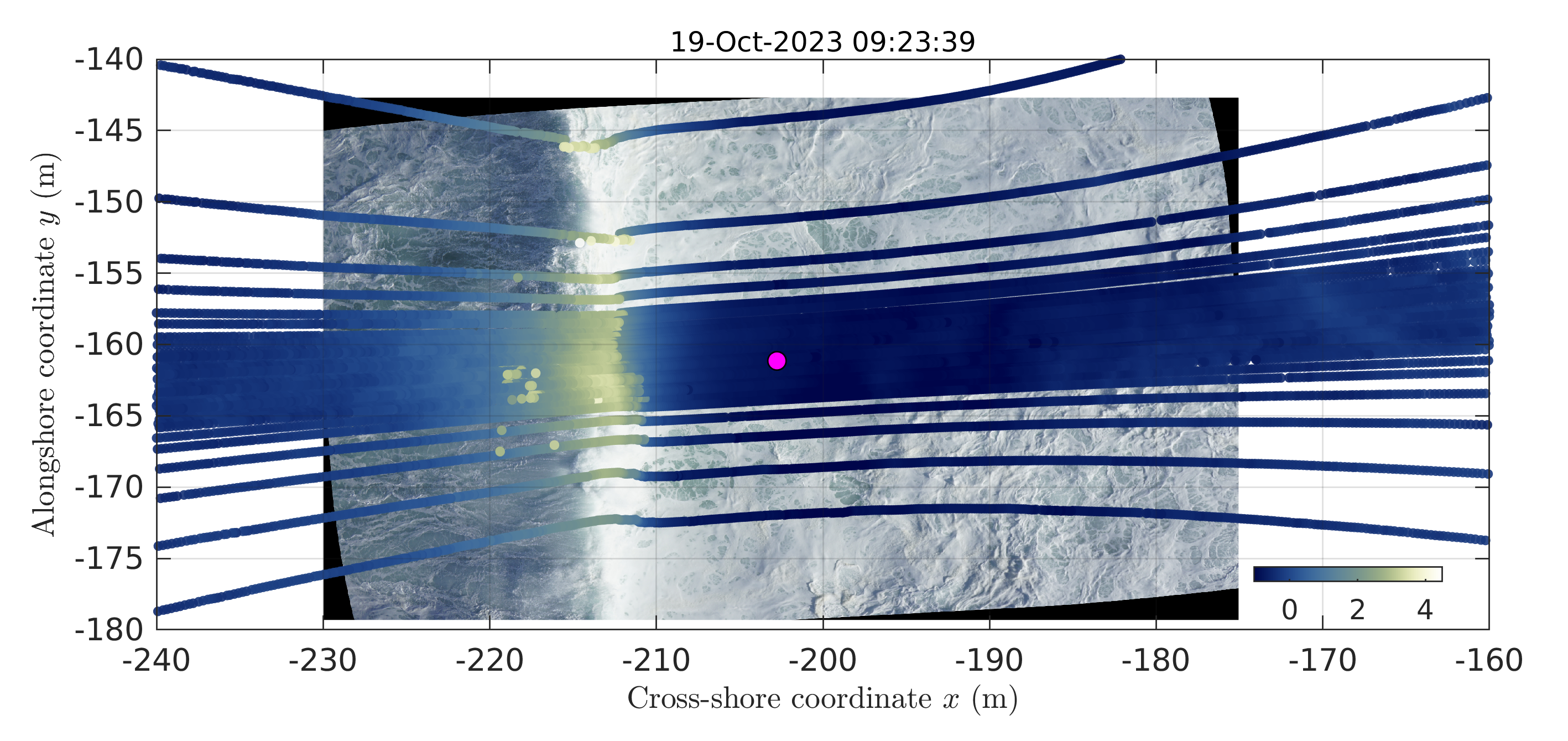}
    \caption{Georectifed sea-surface image in China Rock $(x, y)$ coordinates
      with overlaid lidar-based sea-surface elevation $\eta(x, y)$ (colored
      dots) at 09:23:39 19-October-2023 at time $t=30.0$~s from start of
      lidar data collection.  The magenta circle in Fig.~\ref{fig:map} 
      is the UAS hover location during the capture of these data.  Note the wave crest at $x\approx -219$~m.
    }
    \label{fig:data_plot}
\end{figure}

The lidar data were collected during a large swell event near China Rock (Fig.~\ref{fig:map}) on the coast of the Monterey Peninsula, California, USA.
The overall cross-shore bathymetric slope is (1:40) and the 
shoreline consists of small headlands separated by about 250~m, with 100~m deep embayments.  
The bathymetry is highly variable at scales $<20$~m for depths $h<20$~m.
A local cross- and alongshore  coordinate system $(x,y)$ is defined at China
Rock, with $-x$ direction aligned with $280^\circ$N (Fig.~\ref{fig:map}).

The lidar data were collected during a hover on the 19th of October 2023 at
09:23:39 PST for a total of 701~s.  Here, only the first 60~s are used. The UAS
hovered at $(x,y) = (-204, -162)$~m at 33-m elevation relative to the sea-surface offshore of a headland location
(magenta dot, Fig.~\ref{fig:map}) where large wave breaking occurred.
At this location, the mean water depth (including the tide) was  $h=5.4$~m that increases (decreases)
offshore (onshore) with highly variable rough bathymetric features in the region.  
The winds were light at a few meters per second.
During the hover, a relatively large and building groundswell was present.
A wave buoy 800~m offshore (yellow dot in Fig.~\ref{fig:map}) recorded a sea-swell significant wave height of $H_s = 2.4$~m, a peak frequency of $f_p=0.05$~Hz, and 
an energy-weighted mean wave angle of $\bar{\theta} = -3.2^\circ$.

During the hover, the UAS was oriented such that the lidar scanned in the
cross-shore direction (Fig.~\ref{fig:data_plot}). 
Lidar returns were converted to China Rock $(x,y)$ coordinates.
During the hover the UAS remained at a nearly constant position
with $x$ and $y$ position standard deviations of  $\sigma_x =
0.03$~m and $\sigma_y = 0.02$~m.
The UAS held its orientation consistently during the hover with heading,
pitch, and roll standard deviations $<0.23^\circ$.

At each time $t$ (referenced from the start of the lidar collection), the lidar returns a point cloud $\{(x_i, y_i, z_i, t)
\}_{i=1}^{N_t}$ where $x_i$ and $y_i$ are the cross- and alongshore  China Rock coordinates, 
$z_i$ is the sea-surface elevation, $t$ is time, and $N_t$ is the number of lidar samples at time $t$. 
Note that multiple $x_i$, $y_i$ and $z_i$ correspond to the same time instance $t$ 
and that $N_t \approx 1.5 \times 10^4$ (but the precise number of samples varies with time).
Fig.~\ref{fig:data_plot} shows a lidar point cloud at $t=30.0$~s (referenced from
the start of lidar data collection) along with an RGB image collected at the same time. 
The lidar return density is concentrated in the central region below the UAS and spreads out in the $y$-direction due to the nonlinear distribution of lidar beams (Fig.~\ref{fig:data_plot}).
For the rest of this paper, we consider only the region $-230 \le x \le -180$~m and
$-180 \le y \le -145$~m, where beam returns are relatively dense.
In this region, the lidar captures the elevated sea-surface of the wave crests, consistent with
the RGB image.  Occasional, isolated and elevated noisy lidar returns, likely sea-spray behind the
breaking wave, are also evident (Fig.~\ref{fig:data_plot}).

\subsection{Modeling Lidar Data by SIREN Neural Networks} 
\label{sec:SIREN_reconstruction}
Neural networks (NN) are functions, parameterized by network parameters $\boldsymbol{\theta}\in \mathbb{R}^{n_{\theta}}$, that map inputs $\mathbf{x} \in \mathbb{R}^{n_x}$ to outputs $\boldsymbol{\Phi}(\mathbf{x};\boldsymbol{\theta}) \in \mathbb{R}^{n_{\Phi}}$.
A feedforward neural network (FNN) has the structure of a multilayer perceptron, and is defined by a sequence of
functions $\boldsymbol{\phi}_i : \mathbb{R}^{n_{i-1}} \to \mathbb{R}^{n_i}$ such that
\begin{equation}
    \boldsymbol{\Phi}(\mathbf{x}) = \left(\boldsymbol{\phi}_L(\cdot;\boldsymbol{\theta}) \circ \boldsymbol{\phi}_{L-1}(\cdot;\boldsymbol{\theta}) \circ ... \circ \boldsymbol{\phi}_0(\cdot;\boldsymbol{\theta}) \right) (\mathbf{x}),
    \label{eq:FNN}
\end{equation}
where each $\boldsymbol{\phi}_i$ is given by
\begin{equation}
    \boldsymbol{\phi_i} (\mathbf{x};\boldsymbol{\theta}) = \sigma_i\left(\mathbf{W}_i \mathbf{x} +\mathbf{b}_i\right).
    \label{eq:NN_eq1}
\end{equation}
Here, $\mathbf{W}_{i}$ is an
$n_{i} \times n_{i-1}$ weight matrix and $\mathbf{b}_i$ is a $n_i$-dimensional vector of biases; 
the weights and biases collectively define the network parameters $\boldsymbol{\theta}$;
$\sigma_i(\cdot)$ is the (nonlinear) activation function of the $i$-th layer, which acts element-wise on its input. 
We use a SIREN (sinusoidal representation network) architecture so that $\sigma_i(\mathbf{\gamma}) = \sin( \mathbf{\gamma})$ for all but the output layer (no output activation).
SIRENs have been shown to be efficient, consistent and accurate in regression tasks \citep{sitzmann_implicit_2020}.

A SIREN representation of the sea-surface elevation with network parameters $\boldsymbol{\theta}$ takes in a location and a time, $(x,y,t)$, and outputs an approximate sea-surface elevation, denoted by $\mod$ for simplicity.
We find appropriate network parameters via ``training,'' i.e., minimizing the mean squared error (MSE) loss
\begin{equation}
    L(\boldsymbol{\theta}) = \frac{1}{N} \sum_{j=1}^{N}  \left( z_j(x_j,y_j,t_j) - \hat{\eta}(x_j, y_j, t_j| \boldsymbol{\theta}) \right)^2.
    \label{eq:MSE}
\end{equation}
using the Adam optimizer \citep{kingma_adam_2017} with mini-batching, a popular variant of stochastic gradient descent,
along with cosine annealing~\citep{loshchilov_sgdr_2017} for stepsize control~\citep{sitzmann_implicit_2020}.
As is typical, we normalize the sea-surface elevation data $z_j(x_j,y_j,t_j)$ to have zero mean and unit variance. This transformation is subsequently undone when evaluating the SIREN for sea-surface elevation.
Since the loss~\eqref{eq:MSE} is not convex, the initialization of the network parameters defines the local minimizer found by the numerical optimization and we use ``Xavier initialization,''
which can be used to initialize FNNs whose activation functions satisfy $f(0)= 0$ and $f'(0) = 1$~\citep{glorot_understanding_2010}.
\cite{sitzmann_implicit_2020} derived a different initialization under the assumption that the input data are uniformly distributed.
Lidar data, however, are not uniform, which is why we use Xavier initialization throughout (see also  Section~\ref{sec:disc}).
We implement the optimization in PyTorch and use NVIDIA GeForce GTX 1650 Ti and NVIDIA GeForce GT 1030 graphics cards for computations.

Using the above procedure, we modeled the first 60~s of lidar data (Section~\ref{sec:Methods}\ref{sec:exp_over}) 
with a 4 layer SIREN (300, 500, 500, and 300 neurons -- 551,600 total parameters). 
During training, we used every other frame of lidar data, effectively using a 5~Hz sampling frequency.
At roughly 15,000 data points per frame, there are approximately 4.2 million data points over the 60~s interval.
We split the data into a 80\% training set $\etatrain$, used for optimization, and 20\% testing set $\etatest$ (randomly sampled). 
An 80/20 split for training\slash testing is a common choice \citep{goodfellow_deep_2016}, however the 
lidar observations are inhomogeneously distributed in space, so that testing points concentrate in densely sampled regions,
which is a potential limitation of this approach.  
Training consisted of performing 600 iterations with the Adam optimizer, starting with a learning rate $\alpha = 10^{-3}$ and subsequent cosine annealing.
The total training time on our (outdated) hardware is about $30$ mins.
The final MSE on the training data is $0.018$~m$^2$, corresponding to a root mean square error of about 13~cm.

In computer graphics, SIRENs have been used for
compression of video files, where the size of a parameter set 
defining the SIREN corresponds to the compression of the video 
\citep[\eg][]{kwan_enhanced_2026,rho_neural_2022,zhang2021implicitneuralvideocompression};
efficient point cloud compression has also been developed with SIREN models \citep[\eg][]{zhang_efficient_2025}.
For our problem of reconstructing the sea-surface from lidar data, the compression is also quite strong.
The 60~s of data are a collection of about 4.5 million points,
but the SIREN only requires storing 551,600 network parameters (although we did not optimize the number of layers or layer width), implying a $87\%$ data compression.

\section{Results}
\label{sec:Results}
\subsection{SIREN Reconstructions of the Sea-Surface}
\label{sec:errors}

\begin{figure}[tb]
    \centering
    \includegraphics[width = \textwidth]{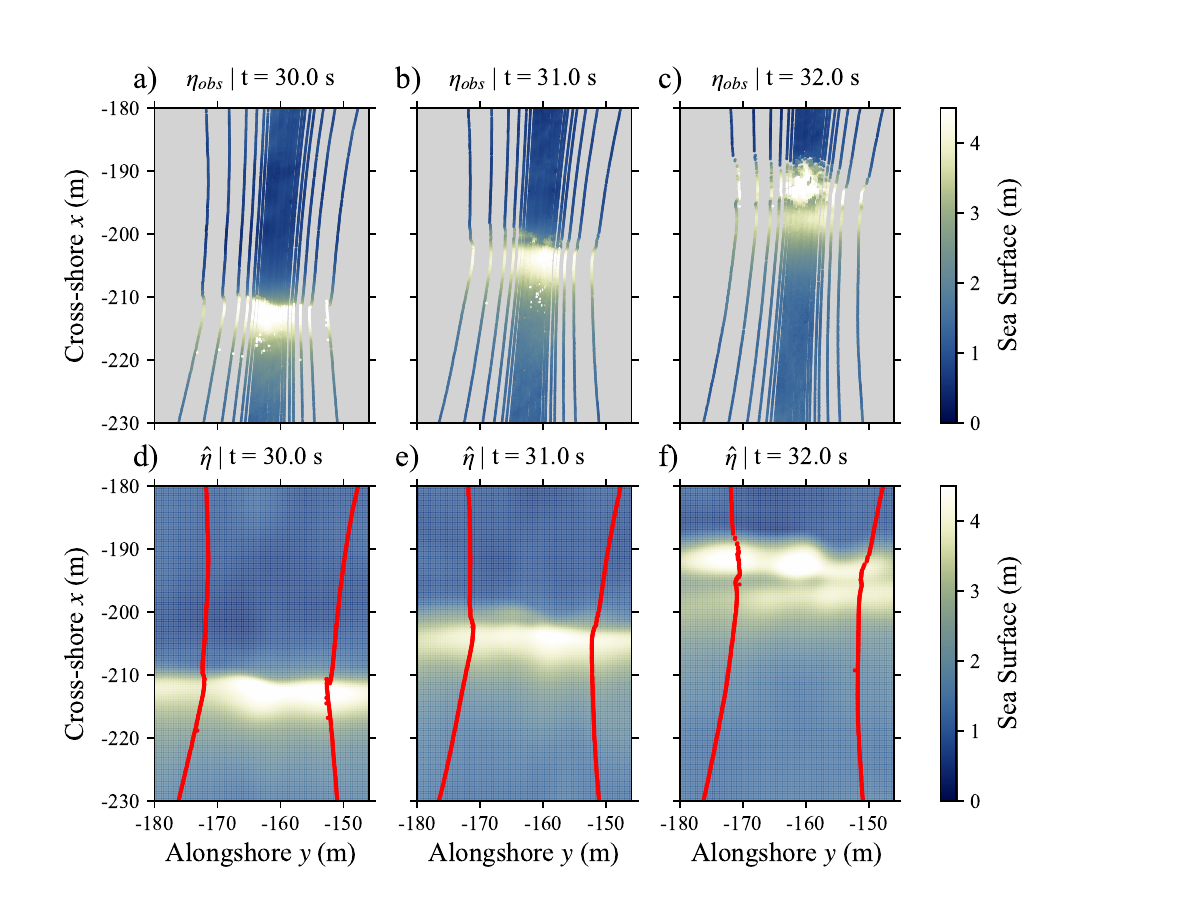}
    \caption{
    (top) Lidar observed sea-surface $\etaobs$ in the  China Rock $(x,y)$ coordinates at times
   (a) $t= 30.0$~s, (b) 31.0~s, and (c) 32.0~s after data collection began. (bottom)
    The SIREN reconstructed sea-surface $\hat{\eta}(x,y,t|\theta)$  (d) $t=30.0$~s, (e)  $31.0$~s, and (f)  $32.0$~s.
    The $(x,y)$ mesh resolution is 0.1~m and outermost lidar observed sea-surface is overlaid in red.}  
    \label{fig:nn_v_data}
\end{figure}

We present the lidar observed sea-surface $\etaobs$ evolution
for two seconds of a strong wave breaking event and compare it to the 
SIREN reconstruction $\mod$, which we evaluate on a regular mesh, roughly defined to be within the bounds of the lidar beams (Fig.~\ref{fig:nn_v_data}). 
At $t=30.0$~s, the lidar observes a large ($\etaobs>4$~m) wave crest near $x=-210$~m
with a relatively flat trough with $\etaobs \approx 0.5$~m at $x>-205$~m (Fig.~\ref{fig:nn_v_data}a).
One second later, the wave crest is still large ($\etaobs>4$~m) and has
coherently propagated onshore nearly 10~m (Fig.~\ref{fig:nn_v_data}b).
Near $(x,y)=(-200,-168)$~m, a secondary hump of elevated $\etaobs \approx 2.5$~m is
present.  
At $t=32.0$~s, the wave crest has propagated farther in $x$ and appears to have
broken with a distinct large splash-up evident near $x=-192$~m, a local minimum near
$x=-195$~m, and a secondary maximum ($\etaobs \approx 3.5$~m) near $x=-198$~m
(Fig.~\ref{fig:nn_v_data}b).
High wavenumber variability is evident in the lidar data, particularly at
$t=32.0$~s.   Isolated artifacts in the lidar data also are evident, likely related
to sea spray.
The lidar data are very rich, and able to observe wave phenomena that in situ
pressure sensors cannot.  However, as discussed above, to use the
lidar data for analysis, it must be gridded in some form.

The SIREN reconstruction $\mod$ does an excellent job of representing the
major features of the propagating and breaking wave within the bounds of the
lidar beams (Figs.~\ref{fig:nn_v_data}d-f).   At $t=30.0$~s, $\mod$
reproduces the key features of the lidar observations
(Fig.~\ref{fig:nn_v_data}d), with some smoothing and removal of artifacts.   
At $t=31.0$~s (Fig.~\ref{fig:nn_v_data}e), $\mod$ has the
leading secondary hump of elevated sea-surface consistent with lidar observations.
At $t=32.0$~s (Fig.~\ref{fig:nn_v_data}f), $\mod$ represents well the
breakup of the wave into the two maxima seen in the lidar.
Thus, the SIREN reconstruction is able to essentially reject outlier points, to lightly smooth $\etaobs$, 
and, perhaps most importantly, the SIREN reconstruction can be evaluated at any $(x,y,t)$,
not just at the times and location where data were collected.

\begin{figure}[tb]
    \centering
\includegraphics[width=\textwidth]{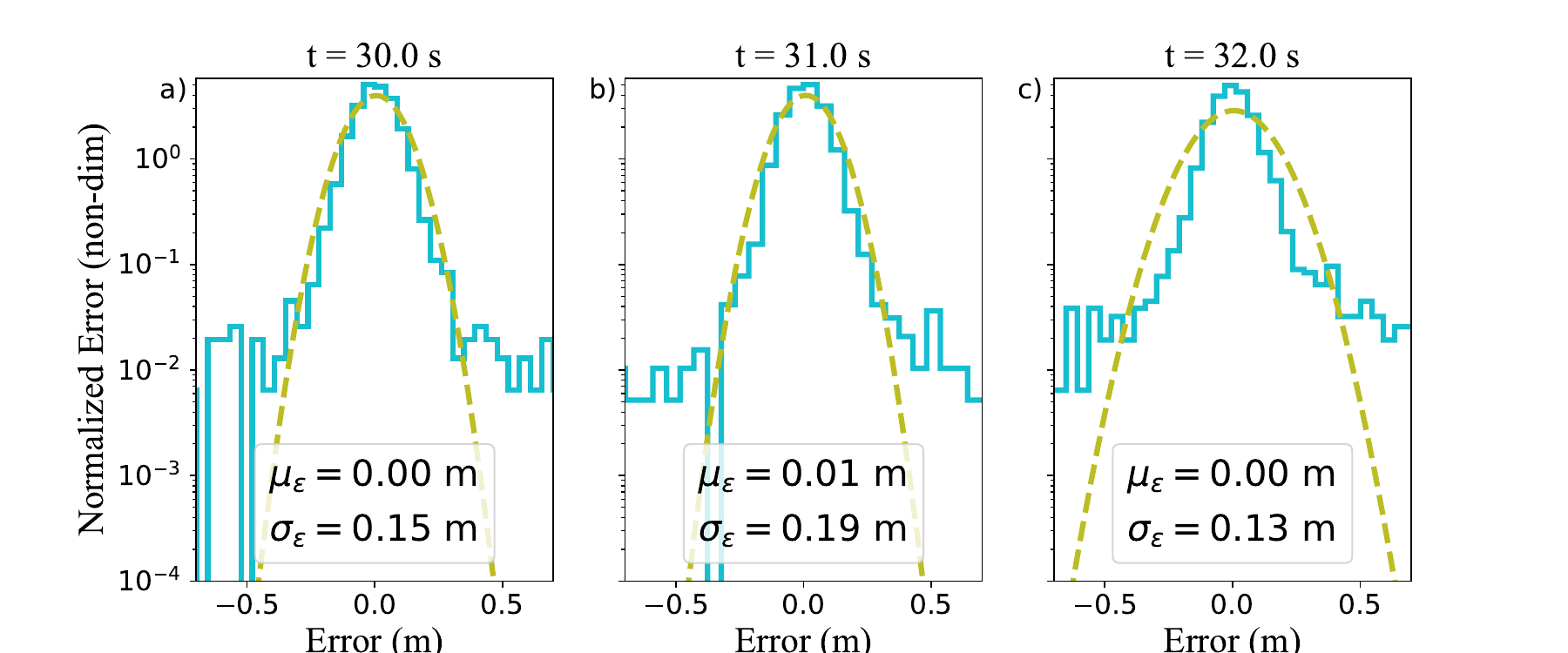}
\caption{Normalized histograms (cyan) of error $\epsilon = \etatest - \heta$ 
  from testing set of lidar returns at (a) $t=30.0$~s,  (b) $t=31.0$~s, and (c)
  $t=32.0$~s.   In each panel, a Gaussian distribution is shown (green dashed) with the given error
  mean  ($\mu_{\epsilon}$) and standard deviation ($\sigma_{\epsilon}$). }
\label{fig:error_hist}
\end{figure}

Recall that the SIREN reconstruction, $\mod$, is trained to minimize the MSE with the training data,
but we can also assess its accuracy on the testing data $\etatest$ (because $\mod$ is a continuous function for any $(x,y,t)$)
We calculate the error $\epsilon = \etatest - \heta$ between the SIREN reconstruction and the testing data $\etatest$ at times $t = 30.0$~s, 31.0~s, and $32.0$~s and estimate the error probability density function (pdf), the mean $\mu_\epsilon$, and standard deviation $\sigma_\epsilon$ (Fig.~\ref{fig:error_hist}).
The mean residual $\mu_\epsilon$ varies from 0.00--0.01~m, depending on the time.  
This indicates that $\mod$ has no to very little bias, relative to the observations.
The standard deviation of error $\sigma_\epsilon$ varies from 0.13--0.19~m,
which is comparable to to the RMS error in the \emph{training} data (see Section~\ref{sec:Methods}\ref{sec:SIREN_reconstruction}).
The lidar instrumental errors are spatially variable and its standard deviation was estimated to be $0.06$~m directly below the hovering UAS \citep{feddersen_estimating_2024}, serving as a lower bound on the expected instrument error. 
Thus, the RMS during training, as well as $\sigma_\epsilon$ during testing, is $2$--$3\times$ the estimated lower bound of instrumental error,
suggesting $\mod$ largely follows $\etaobs$ accounting for the noise in $\etaobs$.
The pdf of $\epsilon$ is partially consistent with a Gaussian distribution with $\mu_\epsilon$ and
$\sigma_\epsilon$ at each time step (Fig.~\ref{fig:error_hist}).   
The pdf of $\epsilon$ is also largely symmetric about the near-zero mean, indicating similar
quality of representation for positive or negative errors.
At $t=30.0$~s and $t=31.0$~s, the pdf of $\epsilon$ is
consistent with a Gaussian out to $1.5\sigma_\epsilon$,
however with far more outliers than a Gaussian distribution.
At $t=32.0$~s, the pdf of $\epsilon$ is narrower than a Gaussian
with even more outliers.  This is likely due to the splash-up 
process of the breaking wave (Fig.~\ref{fig:nn_v_data}c) 
being essentially filtered by the SIREN reconstruction.   

\begin{figure}[tb]
\centering
\includegraphics[width = \textwidth]{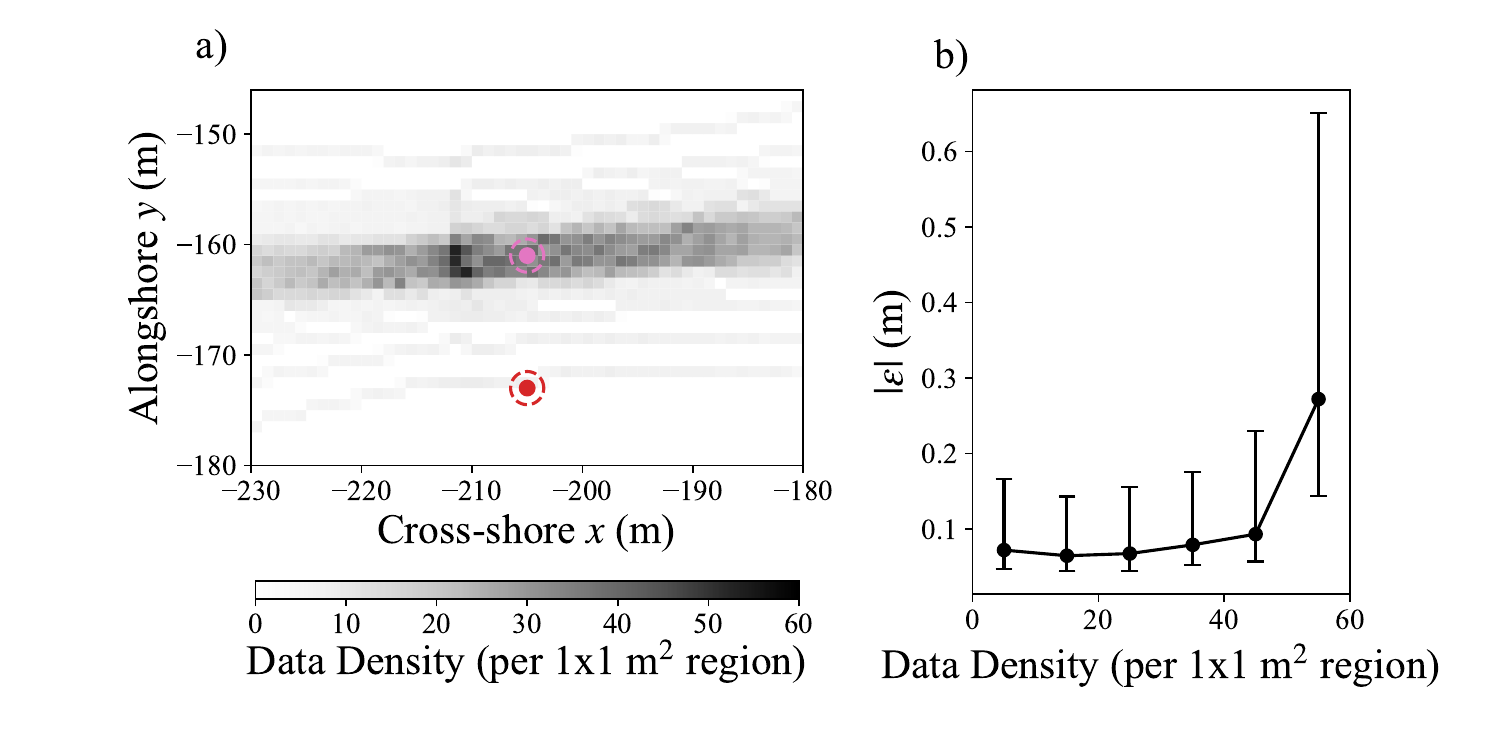}
\caption{(a) 2D histogram  of lidar returns as a function of cross- ($x$)
 and alongshore ($y$) coordinates at $t = 30.0$~s, binned at a 1~m by 1~m resolution.  
 Regions of high data density and low data density are
  indicated as pink and red points surrounded by a circle of radius 1.5~m, 
  respectively.
  (b) 
  Mean (black circle) and 25\%--75\% quartile range (vertical bar) of absolute error $|\epsilon|$, at $t=30.0$~s versus the local data density from (a).  The bin-spacing is such that the number of occurrences in each bin is relatively constant.  
  }
\label{fig:ErrorDensity}
\end{figure}

Lidar data density is variable in the cross-shore $(x)$ and alongshore $(y)$ directions
with the densest region below  the central vertical field of the UAS (Fig.~\ref{fig:data_plot}).
To estimate the density of lidar observations, we binned all data points at a $1\times1$~m$^2$ resolution.
The resulting data density varies from 1 to 50 points per bin (Fig.~\ref{fig:ErrorDensity}a),
where data density is concentrated in the cross-shore direction at the $y$ location of the UAS.
By this definition, any bin without lidar observations has zero data density,
and for any bin with lidar observations the density is $\geq1$. 
The off-center beams typically have between 1-10 data points per bin, and the densest regions under the UAS have 
$>50$ points per bin. 
From this binning procedure, the density of captured lidar is clearly non-uniform which is an important feature for 
using the Xavier initialization, which is further discussed in Section~\ref{sec:disc}.
At data density $\le 35$, the mean $|\epsilon| < 0.1$~m and the 75\% quartile is $<0.2$~m (Fig.~\ref{fig:ErrorDensity}a). 
Even for the sparsest data density, the $\mod$ error is quite low with a mean $|\epsilon|$ of 0.08~m.
This suggests that, in regions of low to moderate data density, the $\mod$ errors are relatively low and quasi consistent with the expected lidar data error.   
However, in regions of high data density ($>35$),
$|\epsilon|$ increases both means and the 25\%--75\% quartile ranges (Fig.~\ref{fig:ErrorDensity}b).
For data density $\geq 55$, the mean $|\epsilon|$ is three times that of the lower data density.
This appears at first counter-intuitive.  
However, the regions of the very highest data density is relatively small
(near $x=-211$~m in Fig.~\ref{fig:ErrorDensity}a) and
includes that of a steep, rapidly varying, and $\approx 4$~m wave crest (Fig.~\ref{fig:nn_v_data}a,d).  
This region is also associated with more sea-spray.
As the SIREN $\mod$ serves to filter high wavenumber variability, 
it will neglect the very high wavenumber variability of sea-spray and somewhat filter the spatial variability of 
the wave crest and overturning wave, particularly as seen at $t=32.0$~s (Fig.~\ref{fig:nn_v_data}c,f).  
Hence, the higher mean error $|\epsilon|$ in the very densest regions
likely is an artifact of $\etaobs$ having high wavenumber variability there.
As most of the reconstructed region is relatively  
flat with low wavenumber variability (Fig.~\ref{fig:nn_v_data}), 
and lower data density regions are evenly distributed across the area of study,
the mean error $|\epsilon|$ is generally lower than in 
the very densest region.

\begin{figure}[tb]
  \centering
  \includegraphics[width=0.95\textwidth]{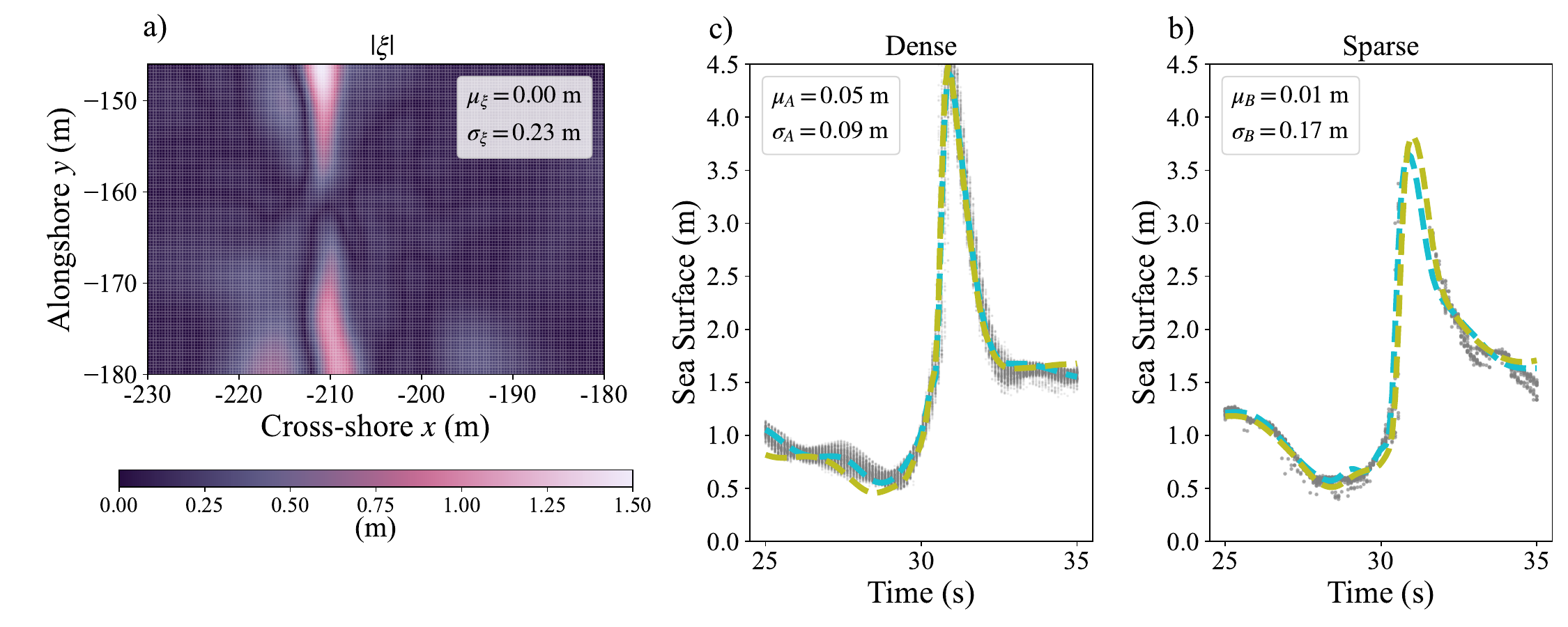}
\caption{ 
 (a) The residuals $\xi(x,y)$ (\ref{eq:xi}) between two SIRENs as a function of $x$ and $y$ at $t=30.0$~s on a fine $(x,y)$ mesh with a grid-size of 0.1~m
 between the two SIREN representations ($\heta_1$ and $\heta_2$), trained on the same data.
 (b,c) $\heta_1$ (cyan) and $\heta_2$ (green) versus time for 
 (c) region of high (magenta dot in Fig.~\ref{fig:ErrorDensity})
  and low (red dot Fig.~\ref{fig:ErrorDensity}) data density at  time resolution of 0.1~s.
Black dots in (b,c) represent lidar returns in the 1.5~m radius circle associated with each dot.
}
\label{fig:analysis}
\end{figure}

\subsection{Reproducibility of SIRENs}
\label{sec:residuals}

During training, we use several random processes -- splitting the data into training and testing sets,
initializing the NN with Xavier initialization, and training using Adam optimization, which is a stochastic method.
To test the reproducibility of our overall approach, we initialize two networks at random (using the Xavier method) and then train both networks on the same training data and using the same number of iterations for optimization.
The result is two sets of optimized network parameters $\boldsymbol{\theta}_1$ and $\boldsymbol{\theta}_2$,
and we denote the corresponding SIRENs by $\hat{\eta}(x,y,t| \boldsymbol{\theta}_1)$ and $\hat{\eta}(x,y,t| \boldsymbol{\theta}_2)$.
The residual between the two models $\xi$ is defined as
\begin{equation}
   \xi(x,y) = \hat{\eta}(x,y,t=30\text{s}| \boldsymbol{\theta}_1) - \hat{\eta}(x,y,t=30\text{s}| \boldsymbol{\theta}_2)
   \label{eq:xi}
\end{equation} 
at $t=30.0$~s  (Fig.~\ref{fig:analysis}a). 
The residuals have a near-zero mean of $\mu_\xi = 0.00$~m, and the standard deviation $\sigma_\xi = 0.23$~m is $3-4 \times$ the estimated lower bound of the instrument error, indicating that $\hat{\eta}(x,y,t| \boldsymbol{\theta}_1)$ and $\hat{\eta}(x,y,t| \boldsymbol{\theta}_1)$ reproduce similar sea-surface elevations on the grid (not just near the data points).  
The absolute residuals, $|\xi(x,y)|$, increase at $x = -210$, which at $t=30.0$~s corresponds to a wave crest of $4$~m,
indicating the increase of residuals is likely caused by the variability in $\etaobs$ at $x =-210$.

Since $\heta_1$ and $\heta_2$ are functions of time, we evaluated them at a point in the dense data region, 
$(x_A,y_A)$ (magenta point in Fig.~{\ref{fig:ErrorDensity}} a), 
and at a point in the sparse data region, $(x_B, y_B)$ (red point in Fig.~{\ref{fig:ErrorDensity}} a), from $t = 25.0$~s to $35.0$~s 
 (Fig.~\ref{fig:analysis}b,c). The residuals are defined by,
\begin{align}
\xi_A(t) &= \heta(x_A, y_A, t \vert \boldsymbol{\theta_1}) - \heta(x_A, y_A, t\vert \boldsymbol{\theta_2})\\
\xi_B(t) &= \heta(x_B, y_B, t\vert \boldsymbol{\theta_1}) - \heta(x_B, y_B, t\vert \boldsymbol{\theta_2})
\end{align}
and we compute the means and standard deviations 
$\mu_A = 0.05$~m, $\mu_B = 0.01$~m, and $\sigma_A = 0.09$~m and $\sigma_B = 0.17$~m.
Both $\mu_A$ and $\mu_B$ are close to zero, suggesting that both SIRENs generate similar sea-surfaces in both the dense and sparse regions.  
The sparse region has a larger standard deviation in the residuals than the dense region,
indicating that both SIRENs can generate variable sea-surfaces in regions where the data density is low. 
The reason is that the training imposes less of a penalty in the data sparse regions.

\begin{figure}[tb]
\centering
\includegraphics[width = .8\textwidth]{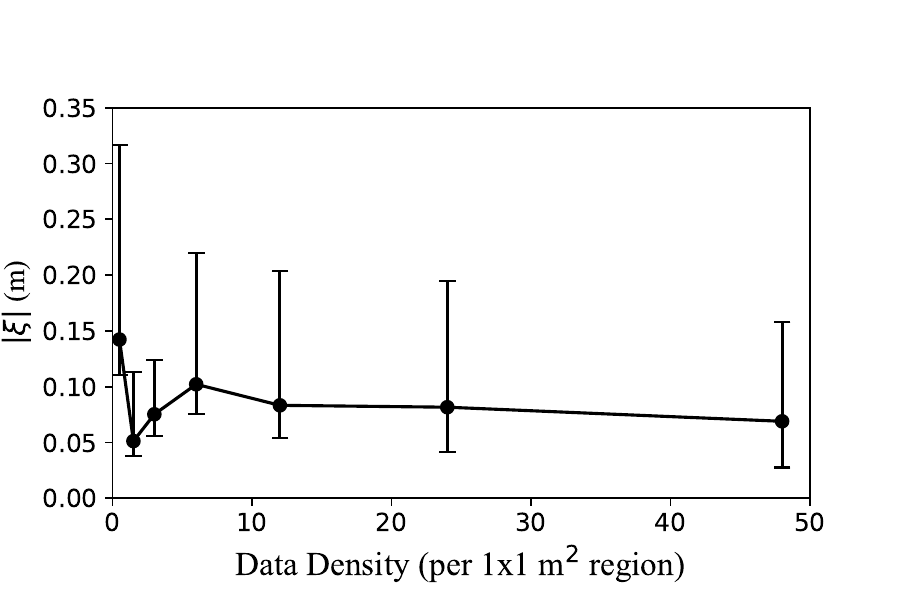}
\caption{Absolute residuals $|\xi(x,y)|$ between two SIRENs at  $t= 30.0$~s versus data density (per $1\times 1~\mathrm{m^2}$) evaluated on a $(x,y)$ mesh resolution of 0.1~m.  
    The black circle represents the mean and the vertical bars represent the 25\%-75\% quartiles.   
    The bin spacing increases geometrically and was set such that the $|\xi(x,y)|$ number in each bin was relatively constant.  }
\label{fig:ResDensity}
\end{figure}

The residuals $\xi(x,y)$ are computed on a mesh over the study region.  Within the region, 
the data density is variable with regions of no data to regions of up to 50 points per $1\times 1~\mathrm{m}^2$. 
We evaluate the dependence of the residual $|\xi(x,y)|$ mean and 25\%-75\% quartiles on the local data density (Fig.~\ref{fig:ResDensity}).  
In regions of sparse data density ($<10$), the mean $|\xi(x,y)| \approx 0.13$~m  with upper 75\% quartile of almost up to 0.40~m.  Thus, in low density regions, 
the difference between SIREN reconstructions becomes relatively large.
In contrast, in regions of high data density ($>20$), the mean $|\xi(x,y)| \approx 0.08$~m, with 75\% quartile up to 0.16~m.   Thus, high data density regions have SIREN reconstructions that are similar to each other. 
The relationship between $|\xi(x,y)|$ and data density is opposite to the relationship between error $|\epsilon|$ and data density (Fig.~\ref{fig:ErrorDensity}). 
The different trends here are likely caused by evaluating the residual between two networks, $|\xi(x,y)|$, in regions with no data (zero data density): Without constraint from data, SIREN reconstructions depend largely on the initialization and, therefore, are somewhat random.  
This should be no surprise since evaluating a SIREN model outside of the training data amounts to extrapolation, which is notoriously difficult to control or do well.
Over all ranges of data density, however, the mean $|\xi(x,y)|$ are all $< 0.20$~m, which is $\times 4$ the estimated instrument lower bound error of $\sigma = 0.06$~m, suggesting that the training procedure overall reproduces similar sea-surface representations within the study region.   

\section{SIRENs for wave analysis}
\label{sec:wave analysis}
We showcase the use of SIRENs for wave analysis via space-time analysis of the sea-surface and by computing spatial and temporal derivatives.

\subsection{Space-time analysis of the sea-surface}
\begin{figure}[tb]
\centering
\includegraphics[width = \textwidth]{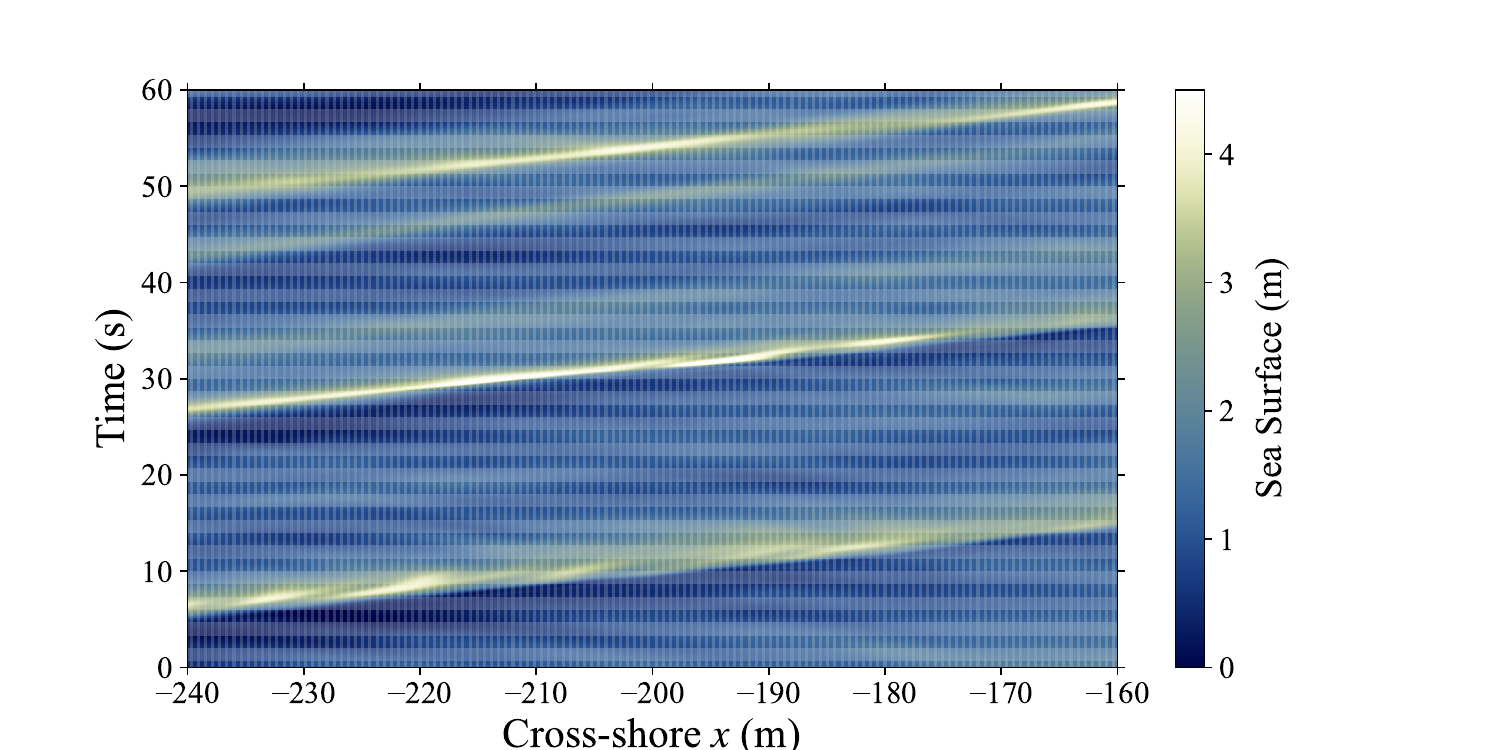}
\caption{The SIREN reconstructed seas surface $\mod$ as a function of $x$ and $t$
  at fixed $y_0 = -162.5$~m.  The $(x,t)$ mesh resolution is $0.1$~m and $0.1$~s.}
\label{fig:hovmoller}
\end{figure}
A classic tool for wave analysis is a Hovm\"{o}ller diagram that illustrates the spatiotemporal variability of a wave field.
Fig.~\ref{fig:hovmoller} shows a SIREN generated Hovm\"{o}ller diagram for the full 60~s time interval and
the 80~m cross-shore range in the center of the alongshore region $y_o=-162.5$~m.
A clear feature are three large wave crests, represented by diagonal bands of elevated ($>4$~m) $\heta$, propagating cross-shore.  
These crests are separated by $\approx 20$~s, which is consistent with the peak period of the wave buoy.   
The 3 bands have similar $x$-$t$ slope that is uniform in the
cross-shore direction, although the bathymetry does get shallower.   
The speed of the wave crest, 
$c = \Delta x/\Delta t$ is estimated from a simple fit of the second wave's crest resulting in $c = 8.75~\mps$.  This is consistent with a nonlinear wave phase
speed of $c_p = (g (h+ H/2))^{1/2} = 8.5~\mps$ using wave height $H=4$~m and mean water
depth $5.4$~m under the UAS.  Thus, the SIREN represents wave propagation and variability well.
The three large wave crests have maximum elevations that increase or decrease
with wave propagation indicating shoaling and wave breaking (Fig.~\ref{fig:hovmoller}).  For example
consider the wave crest between $t=50$~s and $t=60$~s, which reaches a maximum
near $t=55$~s and decreases thereafter with a secondary maximum at $t=59$~s near
$x=-160$~m.  
Two smaller, onshore propagating  waves with weaker crest elevations are also
evident, entering the domain between $t=32$ and $t=43$~s.  These waves propagate
more slowly such that the weak wave crest at $(x,t)=(-240~\mathrm{m},
43~\mathrm{s})$ is close to being captured by the larger crest behind it at $t=60$~s.
Additionally, weak offshore propagating $(x,t)$ bands can be seen behind the
first elevated crest near $(x,t) = (-215~\mathrm{m}, 12~\mathrm{s})$ or
at $(x,t) = (-185~\mathrm{m}, 14~\mathrm{s})$.  These features may be
reflected waves.

\begin{figure}[tb]
\centering
\includegraphics[width=0.8\textwidth]{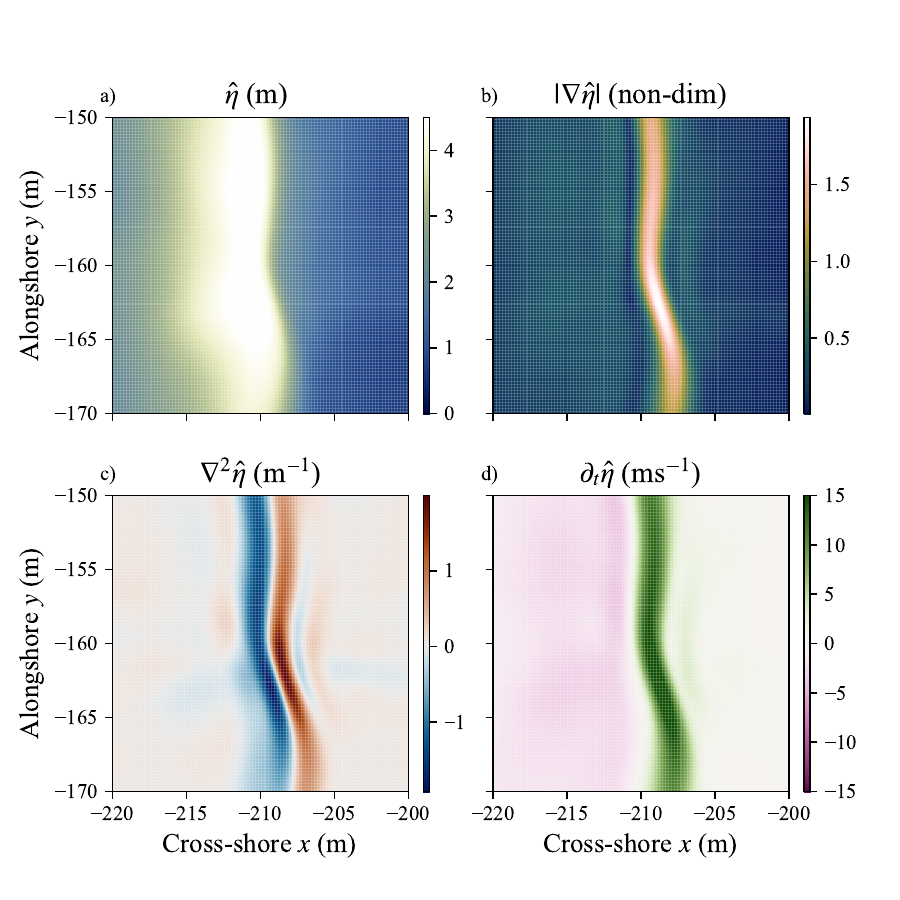}
\vspace{-0.8cm}
\caption{SIREN reconstructed sea-surface and its derivatives at $t=30.0$~s as a
  function of the cross-shore ($x$) and alongshore ($y$) around the wave crest:
  (a) reconstructed sea-surface $\heta$, (b) magnitude of sea-surface gradient
  $\gradeta = ( (\partial_x \heta)^2 + (\partial_y \heta)^2)^{1/2}$, (c)
sea-surface Laplacian $\lapeta  = \partial_x^2 \heta + \partial_y^2
\heta$, and (d) time derivative of sea-surface, $\partial \heta/\partial t$.
 The $(x,y)$ mesh resolution is 0.1~m. }
\label{fig:derivative}
\end{figure}
 
\subsection{Spatial and time derivatives of SIREN  reconstruction} 
Many types of wave analyses, such as directional or nonlinear dispersion
analyses, require estimation of spatial or temporal derivatives of the sea-surface.
Differentiation of NNs is well-understood and automatic differentiation tools allow for efficiently evaluating derivatives of SIREN sea-surface representations at arbitrary locations in space and time \citep{baydin_automatic_2018}.
The SIREN network architecture is particularly stable under differentiation \citep{sitzmann_implicit_2020}.
We now examine various spatial and temporal derivatives of SIREN sea-surface reconstructions $\mod$ at $t=30.0$~s
concentrated at the wave crest (Fig.~\ref{fig:derivative}) to illustrate the differentiability of SIRENs.

As also seen in Fig.~\ref{fig:nn_v_data}e, the SIREN sea-surface $\mod$ is relatively
flat in front of the wave crest ($x>-205$~m), has a steep front-face to a maximum ($\heta >
4$~m), and a back-face with a shallower slope
(Fig.~\ref{fig:derivative}a). The wave crest itself has some alongshore variation.
We estimate $\gradeta  = ( (\partial_x \heta)^2 + (\partial_y \heta)^2)^{1/2}$
using the SIREN $x$ and $y$ derivatives (Fig.~\ref{fig:derivative}b).
In front of the wave crest, $\gradeta$ is small and near $\approx 0.1$.
On the front face, $\gradeta$ has a continuous but bending maximum that varies
from 0.75 to 1.9, suggesting the wave is near overturning.
At the wave crest maximum, $\gradeta$ has a continuous and
bending minimum near zero.
On the back face of the wave, $\gradeta$ is relatively constant in $x$ near a
value of $\approx 0.5$  before flattening out at $x<-220$~m.
Most of the contribution to $\gradeta$ comes from $\partial \heta/\partial x$. 
We can also examine the horizontal Laplacian
$\lapeta  = \partial_x^2 \heta + \partial_y^2$ (Fig.~\ref{fig:derivative}c),
which is related to the curvature of $\heta$. 
Two strong bands of positive and negative $\lapeta$, both with extremal $|\lapeta|
\approx 1.95~\mathrm{m}^{-1}$ are seen on the front face of the wave.
Farther in front of the wave and on the back-face of the wave, $\lapeta$ is weak and
near zero indicating a quasi-planar water surface. 
We can also compute temporal derivatives, e.g., $\detadt$
(Fig.~\ref{fig:derivative}).
On the front face of the wave, a continuous band of $\detadt$ is positive and
large, up to $18~\mps$.
At the wave-crest, $\detadt = 0$, as expected, and also at the same locations where
$\gradeta \approx 0$.
On the back face of the wave, $\detadt$ is much weaker at $\approx -5~\mps$ and
more spatially uniform, also consistent with the $\gradeta$.
The consistent features between $\gradeta$ and $\detadt$ are expected for a
propagating wave.  In fact, for a propagating wave whose shape does not change, $\detadt
= c_p \partial \heta/\partial x$.    Using $c_p \approx 8~\mps$ and maximum
$\gradeta = 1.9$ gives an estimate of $\detadt \approx 15.8~\mps$, close to the
maximum $\detadt = 18~\mps$.   The differences are due to the wave changing
shape and because alongshore wave speeds $c_p$ are neglected.

\section{Discussion and Summary}
\label{sec:disc}
\subsection{Relative merits and limitations of the SIREN representation of lidar data}

SIREN representations of 2D lidar data have advantages and disadvantages
relative to previous nearshore lidar data analysis methodologies.
Most nearshore lidar observations were single beam in the cross-shore direction
with data de-spiked and then binned or interpolated to a fixed grid with
resolution $\Delta x =  0.1$~m
\citep[\eg][]{brodie_lidar_2015,MARTINS2017,Fiedler2021JTECH}.
Conceptually this approach is simple and straightforward, and 
has been used, for example, to estimate wave spectra at various cross-shore locations.
\citep[\eg][]{fiedler_observations_2015,Martins2020JGR}. 
At a $\Delta x = 0.1$~m grid resolution, an extent of 70~m and assuming a 5~Hz sampling frequency over 60~s, 
the single beam gridding uses $2.1\times 10^5$ entries.
The approximately $5.5 \times 10^5$ parameters of our spatially 2D SIREN reconstructed $\heta$ is roughly $2.6$ the 
size of the single beam gridding, while reconstructing a 35~m alongshore extent.   
This indicates that the SIREN representation is quite compact even given its 2D 
nature while still containing significant small scale features (Fig.~\ref{fig:nn_v_data}).

SIREN networks, however are trained on lidar data sets via optimization,
which is more complex and takes longer than simple binning\slash interpolations methods used
previously.
Here, we have only trained on 60~s of data, but computing meaningful waves statistics requires approximating at least 10~min of spatiotemporal lidar data.
Computing a SIREN representation of 10~min of lidar data 
can be achieved by training 10 SIRENs on 10 sequential 60~s windows of lidar data.
Training these 10 SIREN sequentially takes  $10\times 30$~min = 5~hrs, 
but the wall-clock time can be reduced if multiple GPUs are available to train the SIRENs in parallel.
However, training 10 SIRENs over 10 time windows of 60s may produce discontinues.

Training a single SIREN to approximate 10~min of lidar data guarantees continuous sea-surface representations in space and time, but increases compute requirements.
Current GPUs easily accomodate 10~min of lidar data (sampled at 5~Hz) with sufficient memory for mini-batching during the optimization.
More training data generally requires deeper and wider networks (more layers, with more neurons in each layer), but it is difficult to anticipate how much larger the network needs to be to accurately capture 10~min of lidar data.
Some empirical results show that model size (number of total network parameters) grows sub-linearly with data size \citep{hestness2017deeplearningscalingpredictable}.
Assuming a fixed mini-batch size, the total training time can be estimated as follows:
\begin{equation*}
    \text{Training Time for 10 min} \approx \text{Baseline Training Time} \times \text{Data Multiplier} \times \text{Parameter Multiplier},
\end{equation*}
where the baseline training time for 60~s of lidar data is 30~min, the data multiplier is 10 ($10\times$ more data, going from 60~s to 10~min) and where the Parameter Multiplier accounts for the increase in model complexity.
Assuming a $5-10\times$ increase in network complexity then increases the training time to $\sim 1-2$ days.
A training time of 1-2~days is a significant step-up in computational cost compared to gridded approaches, but the spatial and temporal continuity, along with the differentiability of the SIREN representation may be worth this computational investment.

A distinct advantage of the SIREN representation $\heta$ is that it is a continuous
function that can be evaluated, along with derivatives, at any location and at any time \citep{sitzmann_implicit_2020}.
Spatial derivatives can also be estimated from discrete, gridded products,
however, finite difference errors may not be insignificant at steep slopes.
Spatial sea-surface derivatives were accurately estimated at a single
location by fitting a plane or 2D parabola to lidar data within a $O(1)$~m radius 
``instrument''  region \citep{feddersen_estimating_2024}.  
Although it could be done over multiple ``instrument'' regions,
such analysis would be independent.   In contrast, the SIREN representation
$\heta$ is naturally grid-free and allows straightforward estimation of spatial and temporal derivatives. 

Finally, we note that the lidar data we use here have a high point cloud density because the waves
were often breaking and the water was foamy, leading to high returns.   
Farther offshore, the water is clearer and not foamy, and point cloud density may be lower \citep{feddersen_estimating_2024}.
How sea-surface representations via SIRENs would work in regions of lower point cloud density will be a topic for future research.
  
\subsection{Xavier vs SIREN initialization}
\label{sec:xav_sir}

\begin{figure}[tb]
    \centering
    \includegraphics[width = \textwidth]{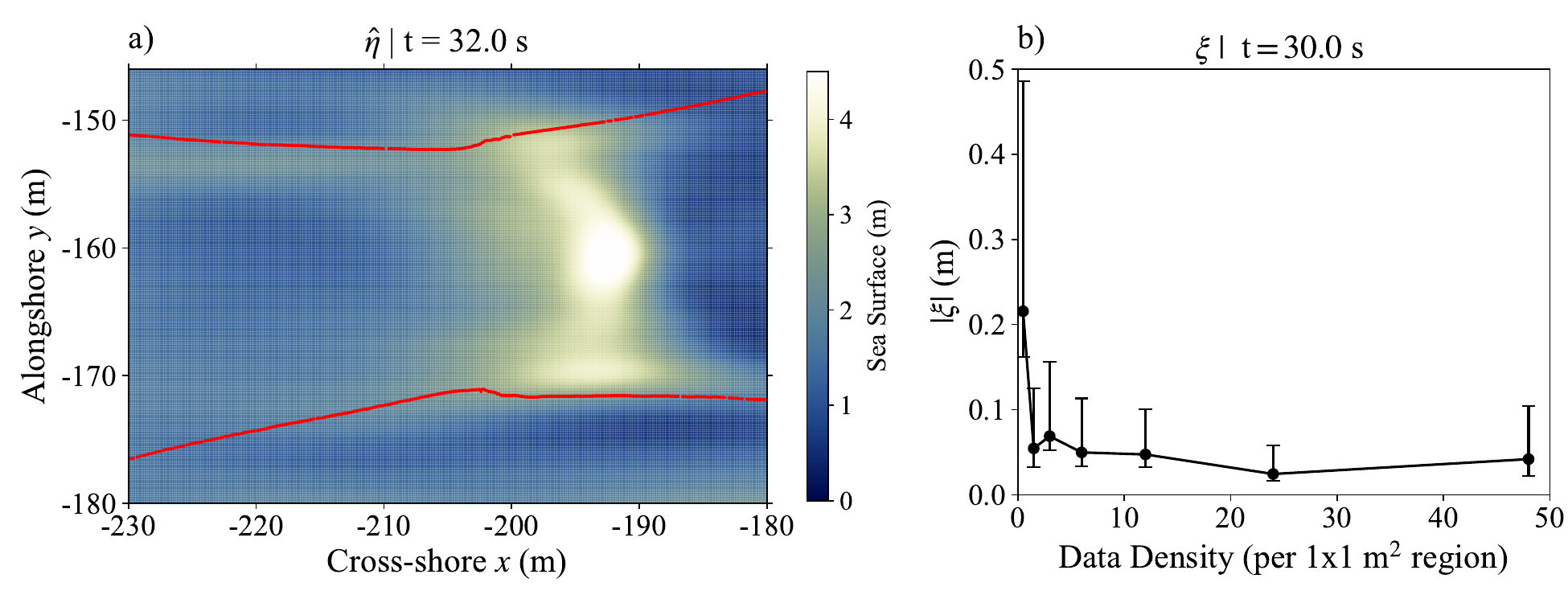}
    \caption{a) A trained SIREN sea-surface at $t=32.0$~s which was initialized with the native SIREN initialization, assuming uniformly distributed data.
     The outermost lidar samples at $t = 32.0$~s are plotted in red.
     b) Absolute residuals $|\xi(x,y)|$ between two SIREN sea-surface representations at $t=30.0$~s which are initialized
     with the native SIREN initialization plotted in the same manner as Fig.~\ref{fig:ResDensity}.}
    \label{fig:init}
\end{figure}

Training a neural network amounts to performing a non-convex optimization over a high-dimensional space with several local minima.
Where we start the optimization, i.e., how we initialize the weights of the network, is therefore an important aspect of the overall training process.
There are several ``standard'' initialization schemes, including Xavier \citep{glorot_understanding_2010} and He \citep{HeInit} initialization.
Which initialization to use depends on the activation function and the overall network architecture.

\citet{sitzmann_implicit_2020} derived a new initialization for SIRENs under the assumption that the input data, after normalization, are uniformly distributed on a unit hyper-cube.
The assumption of uniformly distributed input data is often valid in image or video processing, but this assumption is not valid for the lidar data which is non-uniformly distributed in space.
Each beam samples at a fixed angular rate which results in varying along-beam distribution.  Furthermore, the 32 beams are not evenly distributed in space (Fig.~\ref{fig:data_plot}).

%
%
The fact that the lidar data are not uniformly distributed suggests that the initialization of \cite{sitzmann_implicit_2020} may not be applicable for our purposes.
Indeed, using the ``native'' SIREN initialization on the lidar data results in networks that perform poorly.
For example, we trained a SIREN with the native initialization (Fig.~\ref{fig:init}a). 
At $t=32.0$~s, the network generates a wave with a single wave crest, the sea-surface drops to zero outside of the spatial region where data were available (Fig.~\ref{fig:init}a) with elevated and cross-shore aligned $\heta$ ridges at $-155<y<-150$~m, 
that are not evident in the lidar data or in the SIREN representation we obtained using Xavier initialization (compare Fig.~\ref{fig:nn_v_data}c,f with Fig.~\ref{fig:init}a).
%
The two different initializations report nearly identical testing 
errors, with the native initialization model having an 
MSE of 0.021~m$^2$ and the Xavier initialization model having an MSE of $0.020$~m$^2$.
Since the MSE is a data weighted metric it fails to capture 
how the two initializations vary in sparse regions of data. 
With the native initialization,
the reproducibility statistics were 
worse than with Xavier initialization (compare Fig.~\ref{fig:ResDensity} with Fig.~\ref{fig:init}b).
In regions of high data density, the reproducibility statistics ($|\xi|$) were similar for both initialization schemes.
However, in regions of low data density ($\le 2$), the native SIREN initialization had $|\xi|$ that was 50\% or more larger than that for Xavier initialization.  
%
In summary, the native SIREN initialization results in reduced reproducibility along with poor reconstruction in regions of few\slash no data.

The reason for the improved performance with Xavier initialization (vs. the SIREN-native initialization) is that the Xavier method requires fewer assumptions.
In particular, there is no implicit or explicit assumption about the distribution of the input data (other than that it should have mean zero and finite variance).
The Xavier method for initializing the network parameters may be suboptimal, but for our purposes with non-uniformly distributed inputs, Xavier initialization yielded more accurate SIRENs than using the initialization proposed by \cite{sitzmann_implicit_2020}.

\subsection{Summary}
We developed and tested a method for approximating time sequenced lidar point clouds 
using a specific neural network architecture (SIREN) to create a continuous and differentiable, but grid-free, representation of the sea-surface. 
The point clouds were sampled by equipping a UAS with a multi-beam lidar sensor which
hovered above the surf zone of breaking waves in near-shore rocky conditions off the coast of China Rock, Monterey. 
The lidar sensor projects multiple scans perpendicular to the sea-surface, effectively resolving time stamped
point clouds of the sea-surface. 
A SIREN was then trained over 60~s of the  lidar using the MSE loss, Adam optimization, and the Xavier initialization.
The result is a reconstructed sea-surface that can be queried at any location and at any time and whose derivatives are easy to compute.
The SIREN successfully captures low to medium wave number variability while filtering out the highest wave number variability in 
the sea-surface. 
The errors of the SIREN reconstructed sea-surface are comparable to the instrument errors of the lidar,
and local data density had a minimal effect on the errors found in the reconstruction. 
Training multiple SIRENs over the same data produced sea-surface reconstructions which had residuals comparable to 
instrument error.  We found that in low density regions the residuals tended to be larger than in data dense regions. 
We then evaluated the SIREN at a space-time slice where we estimated wave velocities
consistent with the non-linear wave dispersion.
To further highlight the potential of SIRENs for directional wave analysis, 
we computed spatial and temporal derivatives of the SIREN reconstructed sea-surface.
This method in principle could also be applied to stereo-camera generated point clouds.

\section*{Acknowledgments}

This paper is part of the ROcky Shore: eXperiment and SImulations (ROXSI)
project, funded by the Office of Naval Research through grants N000142112786 supporting FF. 
MM is supported by the US Office of Naval Research (ONR) Grant N00014‐21‐1‐2309.
The Monterey NOAA Sanctuary, CA Fish and Wildlife, and Pebble
Beach provided environmental permission for the experiment
with permitting support provided by Chris Miller from Naval Postgraduate School. We thank the SIO 
field crew for their invaluable support:
Brian Woodward, Kent Smith, Rob Grenzeback, Lucian Parry, Shane Finnerty, Carson
Black. We appreciate the valuable feedback from Jamie H. MacMahan.

\bigskip

\noindent {\bf Data Availability Statement: }
Data and proccesing code can be found on Zenodo at
\href{https://doi.org/10.5281/zenodo.22309467}{https://doi.org/10.5281/zenodo.22309467}

\bibliographystyle{ametsocV6}
\bibliography{references}

\end{document}